\documentclass[prb,showpacs,twocolumn]{revtex4}

\usepackage{graphicx}
\begin{document}

\title{Specific heat study of spin-structural change in pyrochlore Nd$_2$Mo$_2$O$_7$}

\author{Y. Onose$^1$, Y. Taguchi$^2$\footnote{present address: Institute for Materials Research, Tohoku University, Sendai 980-8577, Japan}, T. Ito$^3$, and Y. Tokura$^{1,2,3}$} 
\affiliation{$^1$Spin Superstructure Project, ERATO, Japan Science and Technology Agency (JST), Tsukuba 305-8562, Japan \\$^2$Department of Applied Physics, University of Tokyo, Tokyo 113-8656, Japan \\$^3$Correlated Electron Research Center (CERC), National Institute of Advanced Industrial Science and Technology (AIST), Tsukuba 305-8562, Japan }

%\date{}

\begin{abstract}
By measurements of specific heat, we have investigated the magnetic field ($H$) induced spin-structural change in Nd$_2$Mo$_2$O$_7$ that shows spin-chirality-related magneto-transport phenomena. A broad peak around 2 K caused by the ordering of 2-in 2-out structure of the Nd moments at zero $H$ shifts to the lower temperature ($T$) up to around 3 T and then to the higher $T$ above around 3 T with increasing $H$ for all the direction of $H$. This is due to the crossover from antiferromagnetic to ferromagnetic arrangement between the Nd and Mo moments. While the peak $T$ increases monotonically above 3 T for $H$//[100], another peak emerges around 0.9 K at 12 T for $H$//[111], which is ascribed to the ordering of 3-in 1-out structure. For $H$//[110], a spike like peak is observed at around 3 T, which is caused perhaps by some spin flip transition.
\end{abstract}
\maketitle
Novel magnetic phenomena are frequently observed due to geometrical frustration in pyrochlore oxides where the magnetic ions reside on the vertices of linked tetrahedra. 
One such example is the so-called \lq\lq spin ice\rq\rq  state in $R_2$Ti$_2$O$_7$ ($R$=Ho, Dy).\cite{bramwell} In Dy$_2$Ti$_2$O$_7$, for example, the ferromagnetic interaction among Ising-like Dy moments induces the 2-in 2-out structure, in which the two Dy moments in a tetrahedron point inward and the other two moments point outward. However, there are macroscopically large numbers of spin structures satisfying the 2-in 2-out rule. 
Consequently, the macroscopically degenerate ground state is realized in this material as shown by Ramirez $et$ $al$.\cite{ramirez}

 In Nd$_2$Mo$_2$O$_7$, the Nd moments show the 2-in 2-out structure similarly to the spin ice materials.\cite{taguchi1,yasui} In this material, however, the spin-polarized itinerant Mo 4$d$ electrons coexist with the Nd moments. The Mo and Nd sublattices have the same structure, but are displaced from each other by half a unit cell.
Although the anisotropy of the Mo $4d$ moments is small, the Mo moments are slightly (at most by several degrees\cite{taguchi1,yasui}) tilted from the direction of the net magnetization due to the antiferromagnetic interaction with the Ising-like Nd moments. 
Recently, an unusual behavior of anomalous Hall effect was observed in Nd$_2$Mo$_2$O$_7$ and the origin was proposed to be the spin chirality induced by such a non-coplanar spin structure of Mo moments.\cite{taguchi1}
In this paper, we investigate the spin-structural change in Nd$_2$Mo$_2$O$_7$ by means of specific heat measurements. 
Most of the arguments are concerning the Ising-like Nd moments. We find many behaviors similar to those of the spin ice materials although zero point entropy is not observed in this material. Some of these results provide a supporting evidence of the spin chirality scenario for the unusual anomalous Hall effect.

%Experimental+Mo basic properties
\begin{figure}
\begin{center}
\includegraphics*[width=7cm]{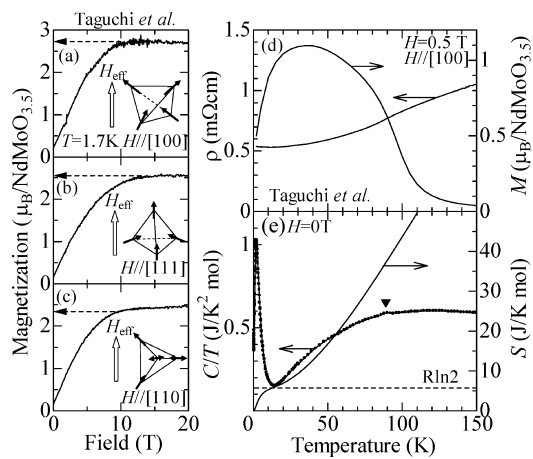}
\caption{(a)-(c): Magnetization curves at 1.7 K with magnetic field ($H$) along (a) [100], (b) [111], and (c) [110] axes. Insets in (a)-(c) show spin structures realized in the high field region of the respective configuration.
(d): Temperature ($T$) dependence of magnetization and resistivity at 0.5 T ($H$//[100]). (e): $T$-variation of specific heat and entropy at 0 T. Closed triangle indicates the ferromagnetic transition $T$ of Mo spin. The magnetization and resistivity data in (a)-(d) are taken from refs. 3 and 5.}
\end{center}
\end{figure}

A single crystal of Nd$_2$Mo$_2$O$_7$ was grown by floating zone method. The details of the sample growth are reported elsewhere.\cite{taguchi1,taguchi2} We reproduce the results of the magnetization and the resistivity for the Nd$_2$Mo$_2$O$_7$ crystal at $H$=0.5 T applied parallel to the [100] direction\cite{taguchi1} in Fig. 1 (d). The resistivity shows a metallic behavior in the whole temperature ($T$) region (2 K $\leq T \leq$ 150 K). The magnetization begins to increase rapidly with decreasing $T$ below around 100 K, reflecting the ferromagnetic transition of the Mo spin ($T_{\rm C} \approx$ 90 K). At around 30 K, the magnetization shows the down-turn due to the ferrimagnetic ordering of Nd moments.
We reproduce the magnetization curves at 1.7 K for the $H$// [100], [111], and [110] axes in Figs. 1 (a), (b), and (c), respectively.\cite{taguchi2} 
For all the field directions, enough high-$H$ reverses the Nd moments, resulting in the
saturate moments that are much larger than the Mo spin moment($\approx 1.4$ $\mu_{\rm B}$/Mo).
For $H$//[100] and [110], the saturate moments are in accord with those expected for the 2-in 2-out structure of the Nd moments (see the insets  of Figs. 1 (a) and  (c)) with the magnitude being $g_{\rm eff}J \approx 2.3$ $\mu_{\rm B}$. These expected values are indicated by the dashed arrows. 
On the other hand, in the case of $H$//[111], the saturate moment coincides with that expected for the 3-in 1-out structure (see the inset of Fig. 1(b)), which is also indicated by the dashed arrow. In the canonical case of Dy$_2$Ti$_2$O$_7$, the metamagnetic transition from the 2-in 2-out to 3-in 1-out structure is observed as a step-like increase in the magnetization curve for $H$//[111].\cite{fukazawa, sakakibara} 
However, any trace of such a step is hardly observed in Nd$_2$Mo$_2$O$_7$ down to 70 mK.\cite{taguchi2}

We measured the specific heat by the conventional relaxation method. The specific heat of Nd$_2$Mo$_2$O$_7$ at zero $H$ is shown in Fig. 1(e). A small peak is discerned around 90 K as indicated by the closed triangle. 
This is owing to the ferromagnetic transition of the Mo spin.
In the low-$T$ region ($\leq$ 15 K), the specific heat is dominated by a intense peak, which is caused by the ordering of 2-in 2-out structure in the Nd sub-lattice. 
These features observed at zero $H$ have already been reported in literature.\cite{yoshii}
We also plot the $T$-dependence of the entropy $S$ deduced from the $T$-integration of $C$($T$)/$T$.
For the analysis, we assumed the linear relation in $C/T$ below the lowest $T$ ($\approx$ 0.4 K).
The $S$ at 15 K is almost in accord with the value of $R\ln 2$, which is expected as from the degree of freedom of Ising moments. Therefore, the entropy of Ising-like Nd moments is mostly released in the low-$T$ ($\leq 15$ K) region.
\begin{figure}
\begin{center}
\includegraphics*[width=6cm]{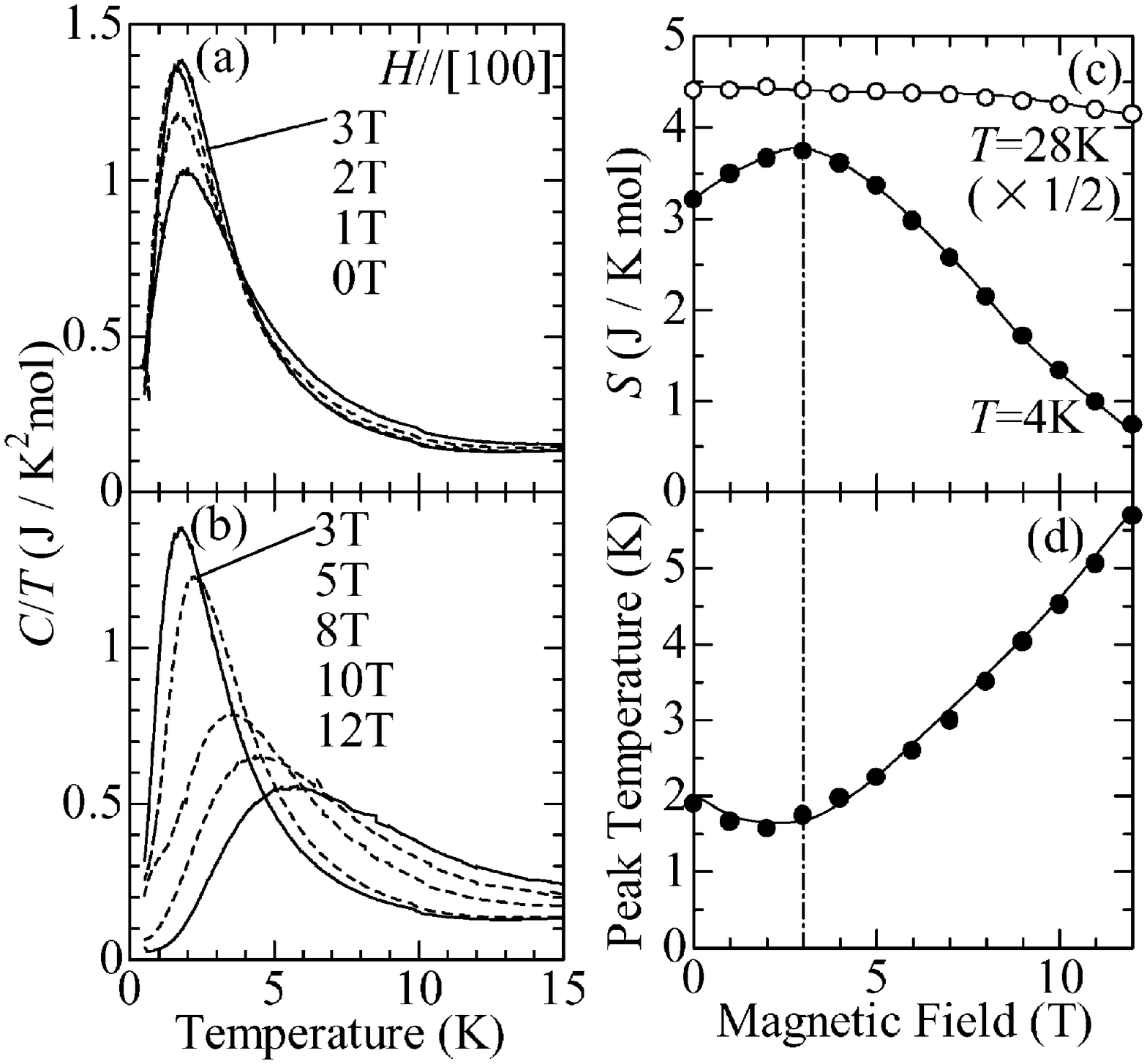}
\caption{(a),(b): Temperature ($T$) and magnetic field ($H$) variation of specific heat for $H//[100]$. (c): $H$-dependence of entropy ($S$) at 4 K and 28 K for $H//[100]$. (d): $H$-variation of the peak $T$ in the specific heat data shown in (a) and (b). The vertical dot-dashed line in (c) and (d) indicates the crossover $H$ from the antiferromagnetic to ferromagnetic arrangement between Nd and Mo moments. Solid lines in (c) and (d) are merely the guide for the eyes.}
\end{center}
\end{figure}

In Figs. 2 (a) and (b), we show the $T$- and $H$-variation of specific heat of the Nd$_2$Mo$_2$O$_7$ crystal below 15 K for $H$//[100], in which all of the four Nd moments in a tetrahedron make the same angle with the $H$ (see the inset of Fig. 1(a)). Whereas the peak due to the ordering of the 2-in 2-out structure becomes sharper with increasing $H$ below 3 T, it becomes broader and the peak $T$ increases as $H$ is increased from 3 T. 
We plot the $H$-variation of the $S$ at 4 K and the peak $T$ in Figs. 2 (c) and (d), respectively. 
Reflecting the crossover around 3 T in the specific heat, the $S$ at 4 K shows down-turn and the peak $T$ shows up-turn around 3 T. 
In the canonical case of Dy$_2$Ti$_2$O$_7$, the ordering $T$ of 2-in 2-out structure increases with increasing $H$ along the [100] direction.\cite{hiroi}
In the present case, the effective field for the Nd moments that is the sum of the applied $H$ and the {\it negative} molecular field from Mo spins changes its sign from negative to positive at around 3 T. This is the reason for the non-monotonic $H$-dependence of $S$ at 4 K and peak $T$. Therefore, the crossover around 3 T can be ascribed to the reversal of the Nd moments, which is consistent with the results of recent neutron measurements.\cite{oohara,yasui2}

We plot the $S$ at 28 K for $H$//[100] in Fig. 2(c). The entropy originating from the  Nd moments is almost completely released below 28 K. 
Almost $H$-independent $S$ at 28 K indicates the absence of zero point entropy even at zero effective field for Nd moments, namely at around $H$=3 T, where the decrease of $S$ would otherwise be observed.
(The slight decrease of the entropy in the high-$H$ region corresponds to its transfer to the higher-$T$ region above 28 K.) Thus, the ground state degeneracy seems to be lifted by the interaction with the itinerant Mo 4$d$ electrons even around 3 T. The $S$ at 28 K shows a similar behavior in the cases of $H$//[111] and $H$//[110] (see Fig. 3(d) and Fig. 4(c)), suggesting that zero point entropy has not been observed in any configuration for Nd$_2$Mo$_2$O$_7$.

%In the high $H$-region, the peak $T$ asymptotically approaches the relation of $T=0.63(H-3)$. In the case of a Schottky-type specific heat for a Zeeman-split of isolated magnetic moments, the peak $T$ obeys the following relation;\begin{eqnarray}
%\label{t-p}
%T_{\rm P}=0.38 \times 2g_{J} J \mu_B H \cos \theta ,
%\end{eqnarray}
%where $T_{\rm P}$ and $\theta$ are the peak $T$ and the angle of the magnetic moment to the $H$. Putting the $T_{\rm P}$/$H_{eff}$=0.63 to eq.\ref{t-p}, we obtain $g_{J} J = 2.1 \mu_B$. This value is similar to the Taguchi $et$ $al$.'s estimation of $g_{J} J \approx 2.3 \mu_B$.

\begin{figure}
\begin{center}
\includegraphics*[width=8cm]{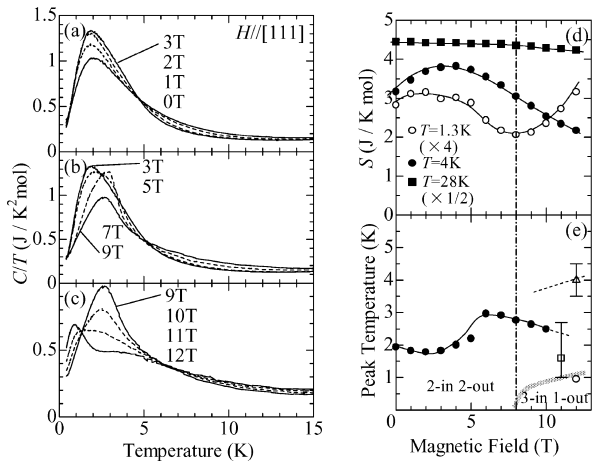}
\caption{(a)-(c): Temperature ($T$) and magnetic field ($H$) variation of specific heat for $H//[111]$. (d): $H$-dependence of entropy ($S$) at 1.3 K, 4 K, and 28 K for $H//[111]$. (e): Characteristic $T$ obtained by the specific heat data for $H//[111]$. Closed and open circles show the peak $T$ due to the ordering of the 2-in 2-out and 3-in 1-out structure, respectively. An open triangle shows the $T$-position of the knee-like structure in the 12 T data. An open square shows the $T$-position of the broader peak in the 11 T data. The vertical dot-dashed line in (d) and (e) represents the crossover $H$ at low $T$ from the 2-in 2-out to the 3-in 1-out structure. Solid, dashed, and hatched lines in (d) and (e) are merely the guide for the eyes.}
\end{center}
\end{figure}

We show the $T$- and $H$-variation of specific heat of the Nd$_2$Mo$_2$O$_7$ crystal for $H$//[111] in Figs. 3(a)-(c). In this configuration, one out of four Nd moments in a tetrahedron is parallel to the $H$ as shown in the inset of Fig. 1(b).
Similarly to the case of $H$//[100], the peak due to the ordering of the 2-in 2-out structure becomes sharper with increasing $H$ up to 3 T owing to the decrease of the total effective field. The crossover around 3 T is also discerned in the $S$ at 4 K as shown in Fig. 3(d).
Above 3 T, the peak becomes broader with increasing $H$ while the shift of the peak is less significant compared with the case of $H$//[100]. Above around 9 T, the specific heat in the low-$T$ region ($T <$ 1.3 K) is gradually enhanced with increasing $H$.
At 11 T, the specific heat shows a much broader peak, which may be viewed as composed of several peaks.
Then, at 12 T, another clear peak emerges at low $T$ ($\approx$ 0.9 K). A knee-like structure is also observed around 4 K in the 12 T data.
It is worth noting here that in Dy$_2$Ti$_2$O$_7$ as the reference material, quite a similar peak emerges at low $T$ when the $H$ (${> \atop \sim}$ 1 T) is applied to the [111] direction.\cite{hiroi} This is ascribed to the ordering of the 3-in 1-out structure (see the inset of Fig. 1(b)). 
In analogy to this, the low-$T$ peak in the 12 T data for the present compound can also be ascribed to the emergence of the 3-in 1-out structure. This assignment is also supported by the fact that the saturate moment for $H$//[111] almost coincides with that expected by the 3-in 1-out structure above 12 T. 
We plot the peak $T$ due to the ordering of the 2-in 2-out structure as with closed circles in Fig. 3(e). 
An upward shift of the peak $T$ is observed around 3 T similarly to the case of $H$//[100]. The peak $T$ shows a kink at 6 T, above which the peak $T$ slightly decreases with $H$.
We also plot in Fig. 3(e) the $T$ positions of the peak due to the ordering of the 3-in 1-out structure and the knee-like structure in the 12 T data with an open circle and triangle, respectively. 
The $T$-position of the broad peak at 11 T is also shown as an open square.
The peak due to the ordering of the 3-in 1-out structure exists in the lower-$T$ region compared with that of the 2-in 2-out structure in the lower-$H$ data. The broad specific-heat peak at 11 T ranging over the wide temperature region (1.0 K$\leq T \leq$ 2.7 K), as indicated by the open square with a long vertical bar, can be ascribed to this heavily mixed state of the 2-in 2-out and 3-in 1-out configuration in this $T$- and $H$-region.

There are several differences in the specific heat for $H$//[111] between the cases of Nd$_2$Mo$_2$O$_7$ and Dy$_2$Ti$_2$O$_7$ while the 3-in 1-out structure is commonly observed in these materials. 
In the specific heat of Dy$_2$Ti$_2$O$_7$, the peak due to the ordering of the Dy moments at the apical position of the tetrahedron (see the inset of Fig. 1(b)) parallel to $H$ emerges in the higher-$T$ region than the ordering $T$ of the 2-in 2-out structure when the $H$ is applied along the [111] direction.\cite{hiroi} This is because the moments parallel to $H$ are more amenable to the $H$ than the other moments. 
However, for Nd$_2$Mo$_2$O$_7$, no clear peak structure of the specific heat is discerned in the higher $T$ region.
Although the specific heat concerning the ordering of the moments parallel to $H$ may be distributed above the peak $T$ of the 2-in 2-out structure in the sufficiently high-$H$ region, the interaction with spin polarized itinerant carriers possibly tends to smear out the high-$T$ peak in the case of Nd$_2$Mo$_2$O$_7$.
(The kink in the peak $T$ around 6 T in Fig. 3(e) may indicate the onset of the separation of the parallel-moment component from the main peak.)
Nevertheless, there is a remnant of such a high-$T$ peak in the specific heat of Nd$_2$Mo$_2$O$_7$, such as a knee-like structure at around 4 K at 12 T.
Another difference is that the spin structure {\it gradually} changes from the 2-in 2-out to 3-in 1-out structure for Nd$_2$Mo$_2$O$_7$ in contrast to the first order phase transition in Dy$_2$Ti$_2$O$_7$. In the specific heat, the low-$T$ component ($T<1.3$ K) gradually increases above around 9 T. These suggest that the correlation of the 3-in 1-out structure evolves even in the lower $H$ region than 12 T. To obtain the onset $H$ of the correlation of the 3-in 1-out structure, we plot the $H$-dependence of the $S$ at 1.3 K in Fig. 3 (d). The quantity shows a minimum at around $H$=8 T, which is thought to be the onset $H$ of the correlation of the 3-in 1-out structure.
 Recently, Taguchi $et$ $al.$ observed the sign change in the Hall resistivity for $H$//[111] and ascribed it to the reversal of the spin chirality caused by the spin-structural change from the 2-in 2-out to 3-in 1-out structure.\cite{taguchi2} The Hall resistivity crosses zero at around $H$=8 T, which coincides with the onset $H$ of the correlation of the 3-in 1-out structure estimated by the present results. This is another evidence supporting the spin chirality scenario for the anomalous Hall effect.

\begin{figure}
\begin{center}
\includegraphics*[width=8cm]{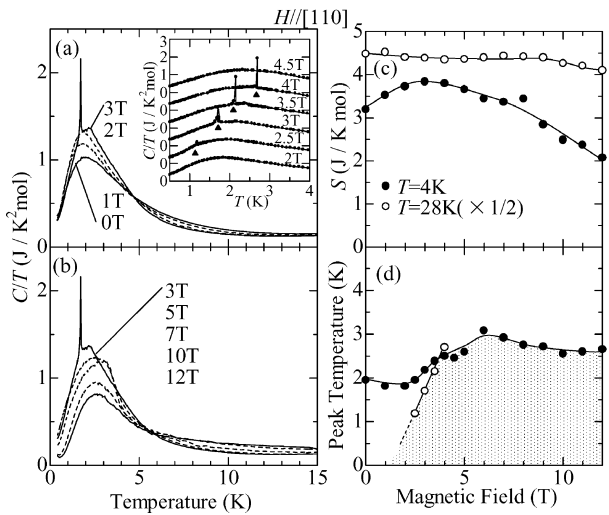}
\caption{(a),(b): Temperature ($T$) and magnetic field ($H$) variation of specific heat for $H//[110]$. Inset in (a) shows the detailed specific-heat data in the $T$- and $H$-region where the spike-like peak is observed. (c): $H$-dependence of entropy ($S$) at 4 K and 28 K for $H//[110]$. (d): $T$-positions of the broad peak around 2 K and the spike-like peak (closed and open circles, respectively). Solid and dashed lines in (c) and (d) are merely the guide for the eyes.} 
\end{center}
\end{figure}

We show the $T$- and $H$-variation of the specific heat of the Nd$_2$Mo$_2$O$_7$ crystal for $H$//[110] in Figs. 4(a) and (b). 
In this configuration, two out of four Nd moments in a tetrahedron are perpendicular to the $H$ (see the inset of Fig. 1(c)).
Similarly to the cases of $H$//[100] and $H$//[111], the peak around 2 K becomes sharper up to 3 T and then broader above 3 T with increasing $H$. The $H$-variation of the peak $T$ and $S$ at 4 K is shown in Figs. 4 (c) and (d), respectively. Corresponding to the crossover around 3 T, the $S$ at 4 K shows the maximum.  The peak $T$ shows a local minimum around 3 T and then a kink around 6 T. In Dy$_2$Ti$_2$O$_7$, a peak emerges in the higher-$T$ region than the peak $T$ due to the ordering of the 2-in 2-out structure.\cite{hiroi2} The higher-$T$ peak is ascribed to the ordering of the half of Dy moments that are not perpendicular to the $H$. In Nd$_2$Mo$_2$O$_7$, by contrast, the higher $T$ peak is smeared out possibly due to the interaction with the Mo $4d$ electrons and the kink around 6 T in the peak $T$ may indicate the onset $H$, above which the half of the moments order ahead of the other two, similarly to the case of $H$//[111]. Because the Nd moments perpendicular to the $H$ are hardly affected by the $H$, the peak $T$ is kept almost constant in the high-$T$ region above 8 T, which is similar to the Dy$_2$Ti$_2$O$_7$ case.\cite{hiroi2}

Another distinct feature for $H$//[110] is the presence of a spike-like peak observed only between 2.5 T and 4.0 T, as exemplified in the inset of Fig. 4(a).
The peak becomes sharper and the peak $T$ increases from 1.2 K to 2.7 K with $H$ in this $H$-region. In Fig. 4 (d), we plot with open circles the $T$ where the spike-like peak emerges. The spike-like peak mainly exists below the peak $T$ due to the ordering of the 2-in 2-out structure. 
Such a spike-like peak as releasing minimal entropy is expected in the case of a spin-flip transition. 
Then, the question is what kind of the spin-flip transition occurs.
One possibility is the aforementioned transition from the antiferromagnetic to ferromagnetic arrangement between Nd and Mo moments. As discussed above, the transition exists around 3 T where the spike-like peak is observed.
However, the transition is not specific for $H$//[110]. Hence, the spike-like peak should be observed also for the other configurations in this case.
Another possibility is the spin-flip transition concerning the Nd moments perpendicular to $H$. Even if the 2-in 2-out structure is assumed, there are two possible arrangements of the Nd moments perpendicular to $H$, namely \lq\lq in-out\rq\rq \ and \lq\lq out-in\rq\rq, in every tetrahedron for $H$//[110] (see the inset of Fig. 1(c)). Therefore, there might be the spin-flip transition between nearly degenerate two spin-structural phases where the arrangements of the perpendicular moments are different from each other.
However, the recent neutron measurement suggests that the net perpendicular moments does not change up to 5 T.\cite{oohara} At present, we cannot draw a definite conclusion about the origin of the spin-flip transition. More detailed investigations in terms of diffraction measurements would be needed.

In summary, we have investigated the specific heat in Nd$_2$Mo$_2$O$_7$ as functions of direction and magnitude of $H$ as well as $T$. A broad peak is observed around 2 K  at zero $H$, which is owing to the ordering of the 2-in 2-out structure of the Nd moments. The peak $T$ decreases up to around 3 T and then increases above around 3 T with increasing $H$ irrespective of the direction of $H$. This is due to the crossover from antiparallel to parallel arrangement between the net magnetizations of the Nd and Mo moments. The peak $T$ increases monotonically above 3 T for $H$//[100].
On the other hand, in the case of $H$//[111], another peak emerges at around 0.9 K at 12 T, owing to the ordering of 3-in 1-out structure of the Nd moments. 
This is consistent with the sign reversal of the Hall resistivity reported by a previous study.\cite{taguchi2}
For $H$//[110], a spike-like peak with minimal entropy release is observed at around 3 T, which is ascribed to some spin flip transition.

The authors thank N. Nagaosa, S. Onoda, H. Takagi, T. Arima, Y. Oohara, and H. Yoshizawa for enlightening discussions. The present work was in part supported by Grant-In-Aid for Scientific Research from the MEXT and by the NEDO.

\end{document}